# Ultraviolet interband plasmonics down to the vacuum UV with ultrathin amorphous silicon nanostructures

**Johann Toudert\*, Rosalía Serna\***
*Laser Processing Group, Instituto de Óptica, CSIC, Serrano 121, 28006 Madrid, Spain*
*\*Corresponding authors*

**Javier Martín Sánchez**
*Department of Physics, University of Oviedo, Spain*
*Center of Research on Nanomaterials and Nanotechnology (CINN), CSIC-University of Oviedo, 33940 El Entrego, Spain*

**Juan I. Larruquert**
*GOLD, Instituto de Óptica, CSIC, Serrano 144, 28006 Madrid, Spain*

**Lorenzo Calvo-Barrio**
*Centres Científics i Tecnològics (CCiTUB), Universitat de Barcelona, C. Lluis Solé i Sabaris 1-3, 08028 Barcelona, Spain*
*IN2UB, Departament d'Enginyeria Electrònica i Biomèdica, Universitat de Barcelona, C. Marti i Franqués, 1, 08028 Barcelona, Spain*

Silicon dominates electronics, optoelectronics, photovoltaics and photonics thanks to its suitable properties, abundance, and well-developed cost-effective manufacturing processes. Recently, crystalline silicon has been demonstrated to be an appealing alternative plasmonic material, both for the infrared where free-carrier plasmons are enabled by heavy doping, and for the ultraviolet where plasmonic effects are induced by interband transitions. Herein, we demonstrate that nanostructured *amorphous silicon* exhibits such so-called interband plasmonic properties in the ultraviolet, as opposed to the expectation that they would only arise in crystalline materials. We report optical plasmon resonances in the 100-to-300 nm wavelength range in *ultrathin nanostructures*. These resonances shift spectrally with the nanostructure shape and the nature of the surrounding matrix, while their field enhancement properties turn from epsilon-near-zero plasmonic to surface plasmonic. We present a vacuum ultraviolet wavelength- and polarization-selective ultrathin film absorber design based on deeply-subwavelength anisotropically-shaped nanostructures. These findings reveal amorphous silicon as a promising material platform for ultracompact and room-temperature-processed ultraviolet plasmonic devices operating down to vacuum ultraviolet wavelengths, for applications including anticounterfeiting, data encryption and storage, sensing and detection. Furthermore, these findings raise a fundamental question on how plasmonics can be based on amorphous nanostructures.

Silicon dominates the electronic, optoelectronic, photovoltaic and photonic industries, thanks to its suitable semiconductor properties, near-infrared bandgap, high visible-to-infrared refractive index, high earthly abundance, and widely developed cost-effective manufacturing processes. Devices based on silicon, such as solar cells, sensors, or electronic, optoelectronic and photonic microchips, are ubiquitous on the market and essential in everyday life.

Recently, with the aim of empowering silicon-based technology using nanophotonic concepts and phenomena, crystalline silicon (c-Si) was unveiled as an appealing alternative plasmonic material, i.e., a plasmonic material beyond the usually studied Drude metals, silver and gold [1]. More specifically, heavy-doping was shown to endow c-Si with Drude plasmonic properties in the mid-to-far infrared (IR) spectral region. These properties were enabled by the increased free charge carrier density induced by doping, and were applied for example to develop integrated molecular sensing applications [2-4].

Very recently, plasmonic properties were revealed across the near and mid ultraviolet (UV) spectral regions in intrinsic c-Si nanostructures [5-14]. Interestingly, in these regions, the plasmonic properties of c-Si are induced by high-oscillator strength interband electronic transitions instead of free-carrier excitations [5,6,14]. Therefore, c-Si was unveiled as a member of the emerging class of so-called "interband plasmonic materials" [14-15], non-Drude plasmonic materials in which the electronic band structure and joint interband electronic density of states drive the plasmonic response, showing great promises for novel functional material platforms intertwining plasmonics, semiconductor physics, band topology, and more.

For interband plasmonic materials in general, and for Si in particular, the crystal structure through the electronic bands is thus expected to markedly affect the material's plasmonic properties. Thus, changes in the atomic order or in the crystal phase are expected to modify the spectral features of plasmon resonances. Therefore, one would expect that amorphous structures *are not able to* support such properties, their weak atomic order precluding high-oscillator strength interband transitions to exist.

However, as opposed to this perhaps naive expectation: *they are*. Herein, we show that a-Si displays interband plasmonic properties from the near UV down to the vacuum UV, and report optical plasmon resonances in the 100-to-300 nm wavelength range in ultrathin nanostructures.

A material is prone to displaying plasmonic properties when the real part $\varepsilon_1$ of its dielectric function is negative. As shown in Figure 1a, which displays the dielectric functions of bulk a-Si, c-Si, Ag, Au, Al taken from the Palik database [16] and of bulk Bi taken from [17], for a-Si $\varepsilon_1 < 0$ across the 94 nm to 317 nm wavelength range. Therefore, a-Si behaves as a plasmonic material from the vacuum UV to the near UV. In this range, upon increasing wavelength, $\varepsilon_1$ first decreases and then increases. After crossing the $\varepsilon_1 = 0$ axis, the $\varepsilon_1$ curve then turns positive and increases until reaching a maximum near 500 nm. This anomalous dispersion behavior results from the Kramers-Kronig consistency of $\varepsilon_1$ with the imaginary part $\varepsilon_2$ of the dielectric function, which shows a broad optical loss band with its maximum near 350 nm. Because this band originates from interband transitions [18-19], a-Si is an interband plasmonic material in the UV.

This behavior is similar to the one of bulk c-Si, as shown in Figure 1a. However, bulk a-Si displays smaller $\varepsilon_2$ values and a broader optical loss band than c-Si. This trend means that, although atomic disorder leads to a broadening in silicon's optical loss features, local order remains adequate to ensure still high-oscillator strength interband transitions in a-Si. Therefore, a-Si displays negative $\varepsilon_1$ values, in a slightly broader UV region than c-Si, and these values are less markedly negative. These trends have an interesting impact on the plasmonic features of a-Si compared with c-Si: respectively (i) they can be achieved in a broader region, (ii) the so-called epsilon near-zero plasmonic properties, which occur when $\varepsilon_1$ is negative and close to 0, can be more robustly achieved.

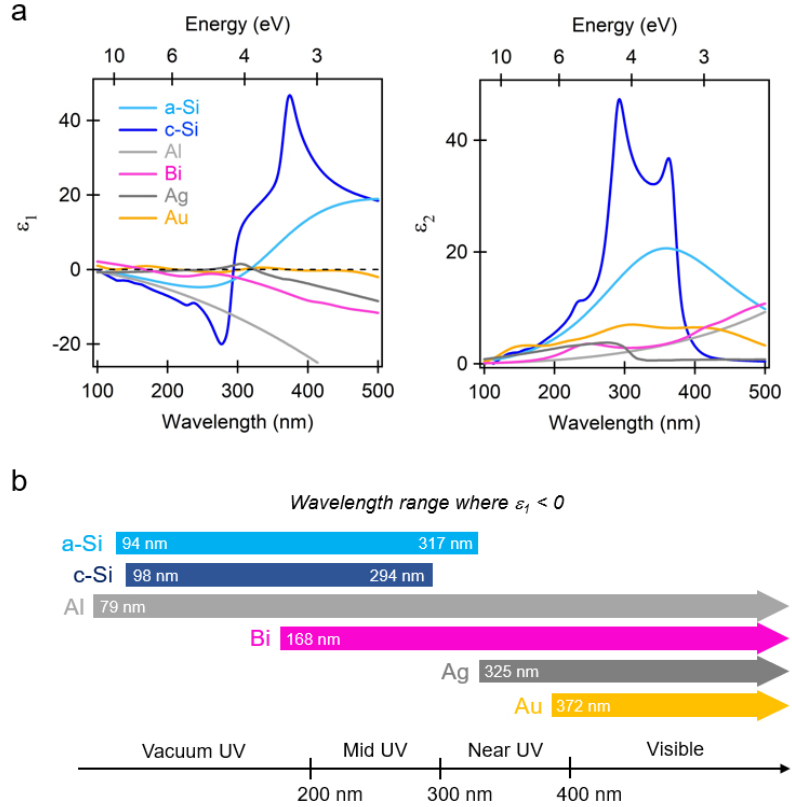

**Figure 1. UV interband plasmonics with a-Si and comparison with other plasmonic materials.** (a) Spectra of the real and imaginary part of the dielectric function $\varepsilon = \varepsilon_1 + i\varepsilon_2$ of bulk a-Si, c-Si, Al, Bi, Ag, Au, in the vacuum UV-to-visible range. All the data were taken from [16] except for Bi, for which they were taken from [17]. (b) Overview of the spectral regions in which these materials display a negative $\varepsilon_1$, where plasmonic effects can occur.

As shown in Figure 1a and Figure 1b, a-Si enables plasmonic properties in the UV while the standard Drude plasmonic materials, Ag and Au, do not. Its UV plasmonic bandwidth overcomes the one of the archetypal interband plasmonic material, Bi [20-21], for which $\varepsilon_1 < 0$ only for wavelengths longer than 168 nm. Furthermore, a-Si competes in the vacuum and mid UV with the most popular UV Drude plasmonic material, aluminum [22-26], the vacuum-to-mid UV plasmonic bandwidth being 94 - 300 nm for a-Si vs 79 - 300 nm for Al, and a-Si showing less negative $\varepsilon_1$ values.

To further demonstrate the UV plasmonic character of a-Si, we calculated the effective dielectric function $\varepsilon_{eff} = (n_{eff} + ik_{eff})^2$ of planar arrays of ultrathin monodisperse nanostructures embedded in a usual wide bandgap amorphous aluminum oxide (a-Al$_2$O$_3$) matrix, as shown in Figure 2a. Calculations were done using a realistic and advanced model [27]. The incident light was unpolarized and impinging at normal incidence onto the array. Such normal incidence conditions were used to achieve all the simulations and experiments results shown hereafter, where we will make an emphasis on the effective extinction coefficient $k_{eff}$ accounting for optical absorption. Since a-Al$_2$O$_3$ displays optical absorption only for wavelengths shorter than 200 nm, optical absorption at longer wavelengths in the $k_{eff}$ spectrum will result exclusively from the a-Si nanostructures.

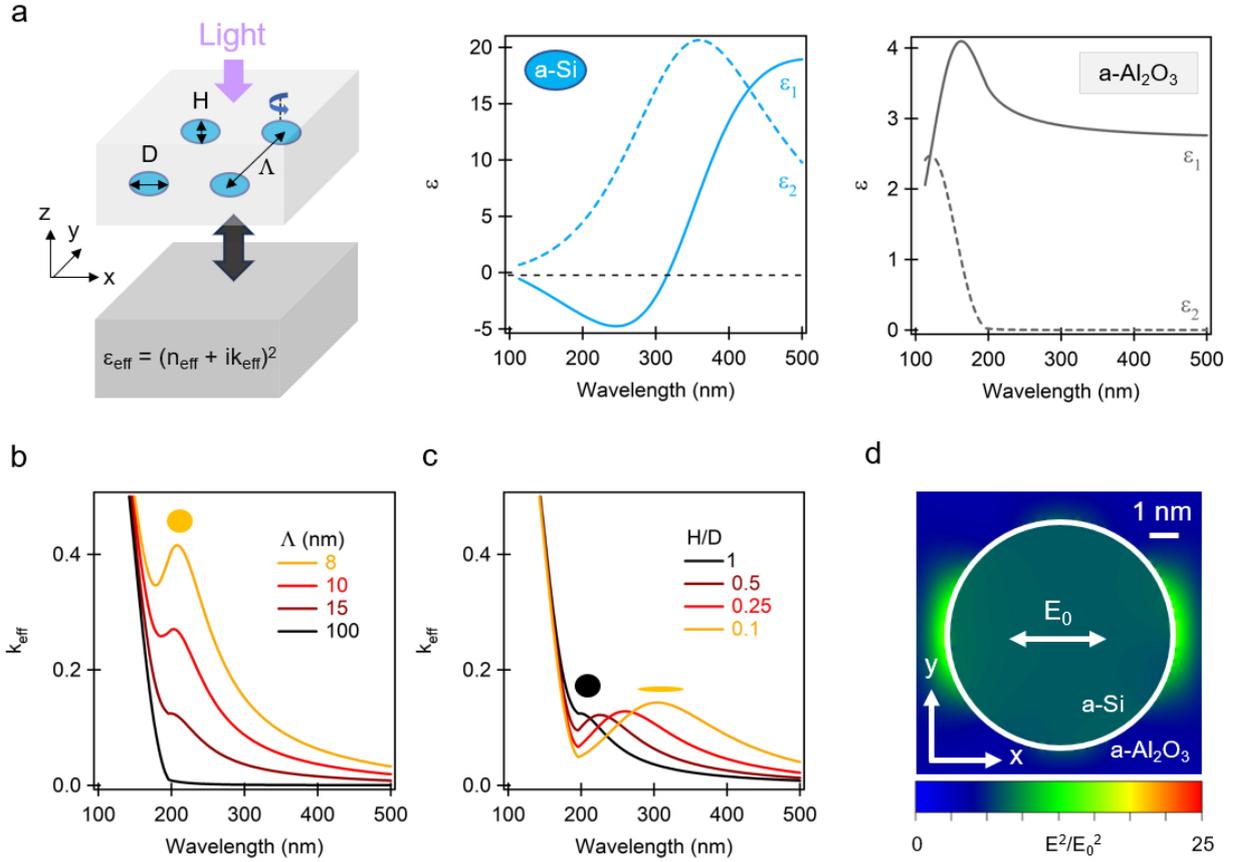

**Figure 2. UV plasmon resonances in ultrathin a-Si nanostructures: simulation.** (a) Left panel: a planar square array of monodisperse spheroidal a-Si nanostructures embedded in a-$Al_2O_3$ and distributed in the x-y plane is modeled as an effective medium. Their revolution axis is oriented along the z direction, their in-plane diameter is D, their height is H, and their separation distance is Λ. Unpolarized light impinges along the z axis, at normal incidence. Center and right panel: dielectric function $\varepsilon = \varepsilon_1 + i\varepsilon_2$ of a-Si and a-$Al_2O_3$ taken from [16] and measured (see Methods section), respectively. Simulated spectra of the effective $k_{eff}$ extinction coefficient for: (b) spherical nanostructures (D = H = 6 nm) as a function of Λ, (c) spheroidal nanostructures (Λ = 15 nm) as a function of the H/D ratio at fixed nanostructure volume. (d) FDTD-calculated electric field map of a 6 nm spherical a-Si nanostructure for the electric field $E_0$ of the incident light oriented along the x axis.

As shown in Figure 2b, the $k_{eff}$ spectra calculated for spherical 6 nm a-Si nanostructures as a function of their separation distance Λ display two main features. At wavelengths shorter than 200 nm, they are dominated by the bandgap absorption tail of a-$Al_2O_3$. At longer wavelengths, they display a resonant absorption band with its maximum near 220 nm, the amplitude of which increases as Λ decreases, i.e., as the areal density of nanostructures increases. As shown in Figure 2c, this band shifts towards longer wavelengths as the aspect ratio H/D of the nanostructures decreases, i.e., as they turn from spheres into flattened, oblate spheroids. This trend is characteristic of localized plasmon resonances in nanostructures. This reveals the plasmonic origin of the observed absorption band, which is further confirmed by the dipolar surface plasmonic-enhanced near-field of the nanostructure shown in Figure 2d.

To experimentally demonstrate that UV plasmon resonances exist in ultrathin a-Si nanostructures, we fabricated a planar assembly of such nanostructures embedded in an a-$Al_2O_3$ matrix. Using the pulsed laser deposition technique, an a-$Al_2O_3$/a-Si/a-$Al_2O_3$ layered film was deposited at room temperature [28]. The amount of a-Si deposited was so small that a discontinuous layer was formed, as shown in the cross-section HRTEM image of Figure 3a. This layer consists of a-Si nanostructures, the height of which is in the 3 nm range. They are elongated in the film's plane, in which they show a polydisperse projected size distribution and are randomly oriented, as usually observed in layers grown in the coalescence-to-near percolation regime. The XPS in-depth spectra presented in Figure 3b confirm the presence of Si-Si bonds and thus of Si nanostructures in $Al_2O_3$, and show the existence of Si-O bonds that likely formed at the surface of the Si nanostructures.

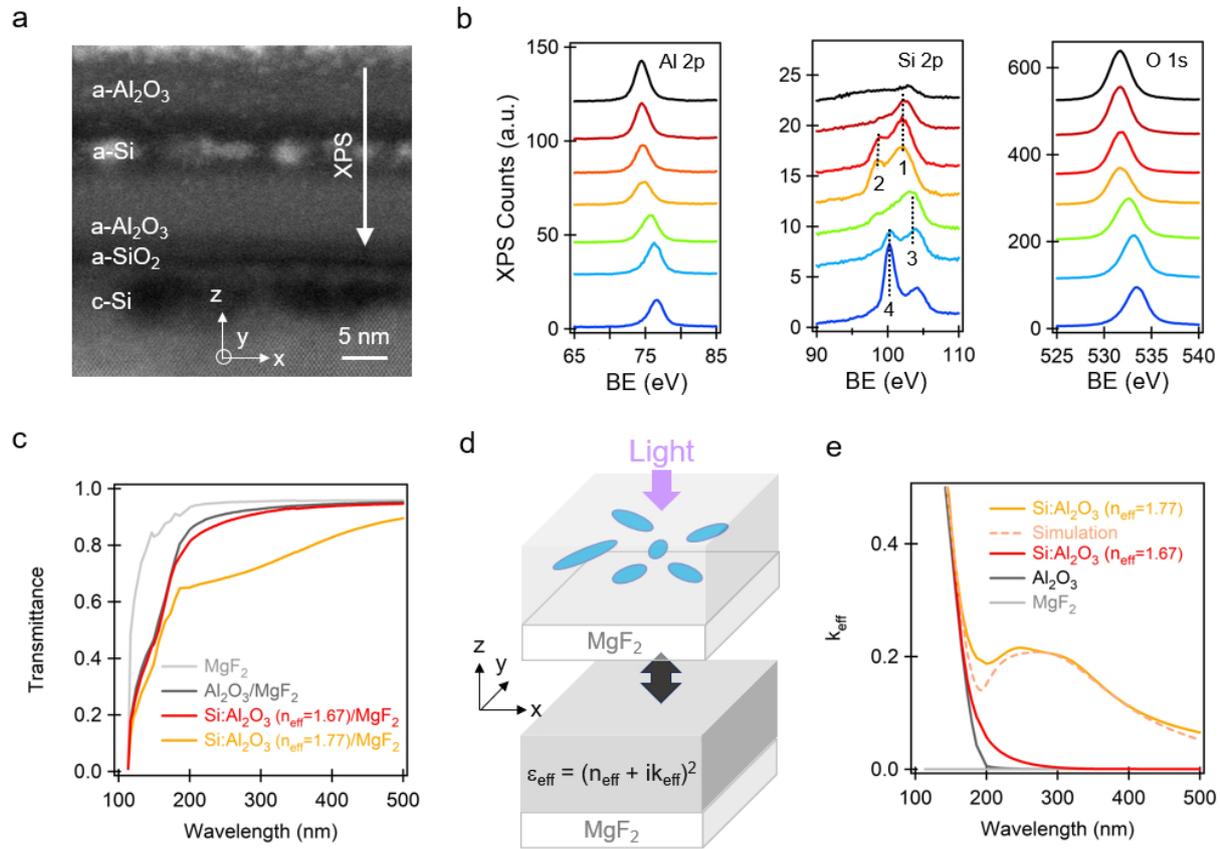

**Figure 3. UV plasmon resonances in ultrathin a-Si nanostructures: experiment.** (a) Cross-sectional HRTEM image of an a-$Al_2O_3$/a-Si/a-$Al_2O_3$ layered film deposited on a c-Si substrate. (b) In-depth XPS spectra of the binding energy (BE) regions for the Al 2p, Si 2p and O 1s orbitals. The Si 2p photoelectron bands account for 1: Si(+iv), atoms at the nanostructure surface; 2: Si(0), atoms in the nanostructures; 3: Si(+iv), native $SiO_2$ at the substrate surface; 4: Si(0), c-Si substrate. This film is labeled Si:$Al_2O_3$ ($n_{eff}$ = 1.77) according to its visible-to-near infrared effective refractive index. (c) Optical transmittance spectra of the Si:$Al_2O_3$ films with $n_{eff}$ values of 1.77 and 1.67, and of the reference a-$Al_2O_3$ film, and of the $MgF_2$ substrate used to deposit the films. Note that the higher the $n_{eff}$ the larger the Si content in the films. Measurements were done with unpolarized light at normal incidence. (d) Each film was modeled as an effective medium to extract its $k_{eff}$ spectrum (see Methods section). (e) Corresponding $k_{eff}$ spectra and spectrum simulated for the Si:$Al_2O_3$ ($n_{eff}$ = 1.77) film assuming ellipsoidal a-Si nanostructures elongated and randomly oriented in the x-y plane.

The optical transmittance spectrum of this film is shown in Figure 3c, together with that of another one containing a lower amount of Si. These films are labeled according to their visible-near IR effective refractive index $n_{eff}$, which reflects their different Si content: Si:$Al_2O_3$ ($n_{eff}$ = 1.77) and Si:$Al_2O_3$ ($n_{eff}$ = 1.67), respectively. The transmittance spectrum of a reference a-$Al_2O_3$ film ($n_{eff}$ = 1.64) is also shown. The total a-$Al_2O_3$ thickness is comparable in the 3 films, which were deposited on UV-transparent $MgF_2$ substrates (see Methods section).

In comparison with $MgF_2$, the Si:$Al_2O_3$ ($n_{eff}$ = 1.77) film shows two main features: (i) a lower transmittance tail in the sub-200 nm wavelength range, (ii) a broad and lower transmittance band spanning over the 200 – 500 nm wavelength range. The spectra of the Si:$Al_2O_3$ ($n_{eff}$ = 1.67) film and reference $Al_2O_3$ film only display the sub-200 nm tail. Moreover, they are very similar, because the a-Si content in the Si:$Al_2O_3$ ($n_{eff}$ = 1.67) film is very low, as accounted by its $n_{eff}$ value close to that of a-$Al_2O_3$ (n= 1.64). Since only the sub-200 nm tail is observed for these two films, it is due to a-$Al_2O_3$. In contrast, the 200 - 500 nm band, which is seen only in the film with the highest a-Si amount, is due to the a-Si nanostructures. Note that this band could by no means be observed if a-Si was formed as a continuous ultrathin layer, as shown in Supporting Information S1, and thus it can be concluded that it results from a discontinuous a-Si layer.

To fully address the physical origin of these spectral features, we extracted the $k_{eff}$ spectra of the 3 films from the measured spectra using transfer matrix modeling and Kramers Kronig-consistent multi-oscillator dielectric functions (see Methods section). As shown in Figure 3d, each film was considered as a homogeneous effective medium [29]. The extracted $k_{eff}$ spectra are shown in Figure 3e, where the extinction spectrum of $MgF_2$ is included for comparison. The sub-200 nm tail and the 200-500 nm band both clearly appear in the $k_{eff}$ spectrum of the Si:$Al_2O_3$ ($n_{eff}$ = 1.77) film, proving that they are both due to optical absorption. Therefore, it is confirmed that the sub-200 nm tail is due to absorption in a-$Al_2O_3$ and the 200-500 nm band is due to absorption in the a-Si nanostructures.

Furthermore, as shown in Figure 3e for the Si:$Al_2O_3$ ($n_{eff}$ = 1.77) film, both this tail and band are well accounted for by effective medium simulation, using the model described in [27]. The simulation assumed a planar assembly of ellipsoidal a-Si nanostructures randomly in-plane oriented and distributed, and embedded in a-$Al_2O_3$. The nanostructures height was H = 3 nm, their in-plane length a = 19 nm, their in-plane width b = 7.6 nm, and their mean separation distance Λ = 15 nm. The corresponding surface coverage is 52%.

Such nanostructure shape and coverage are consistent with a discontinuous a-Si layer in the coalescence-to-near percolation regime, as observed in the cross-section HRTEM image (Figure 3a). Note also that the a-Si volume fraction in the film calculated from the values of a, b, H and Λ is f = 5.7%, in excellent agreement with visible-near IR spectroscopic ellipsometry measurements yielding f = 5.8%. Such agreement with independent measurements together with the measured $k_{eff}$ spectrum points at the suitability of our modeling approach.

From the data discussed above, we definitely address the origin of the 200-500 nm absorption band. It results from plasmon resonances in the a-Si nanostructures. Because of their in-plane elongated shape, 2 plasmon modes can be probed by light at normal incidence, a longitudinal and a transverse one that resonate at different UV wavelengths. Because the nanostructures are randomly in-plane oriented and the incident beam is unpolarized, both modes are excited in a single measurement. They spectrally overlap so that a broad absorption band spanning across the 200-500 nm range is observed.

Therefore, our analysis demonstrates that UV interband plasmon resonances take place in ultrathin a-Si nanostructures embedded in a matrix (in this case, a-$Al_2O_3$). In Supporting Information S2, evidence of such resonances is given in other films containing different amounts of a-Si. It is shown how they shift and broaden toward longer wavelengths as the a-Si content increases. This trend, which results from a decrease in the nanostructure's aspect ratio as shown in Figure 2c, is similar to that observed for plasmon resonances in nanostructures based on other materials, such as Ag, Au, Al or Bi [20,23,27].

Beyond this fundamental understanding, the reported analysis enables finding design concepts for ultrathin films fully harnessing the plasmonic spectral region of a-Si, i.e. from 94 to 317 nm. Indeed, the films shown above yield observable plasmonic effects only at wavelengths longer than 200 nm, so that alternative designs are required to grant access to the sub-200 nm, vacuum UV region. To approach such designs, we performed FDTD simulations of the optical transmittance spectra of planar square arrays of monodisperse a-Si nanostructures embedded in a matrix with dielectric function $\varepsilon_m$.

The array is shown in Figure 4a. In contrast with previous simulations, the incident light was linearly polarized.

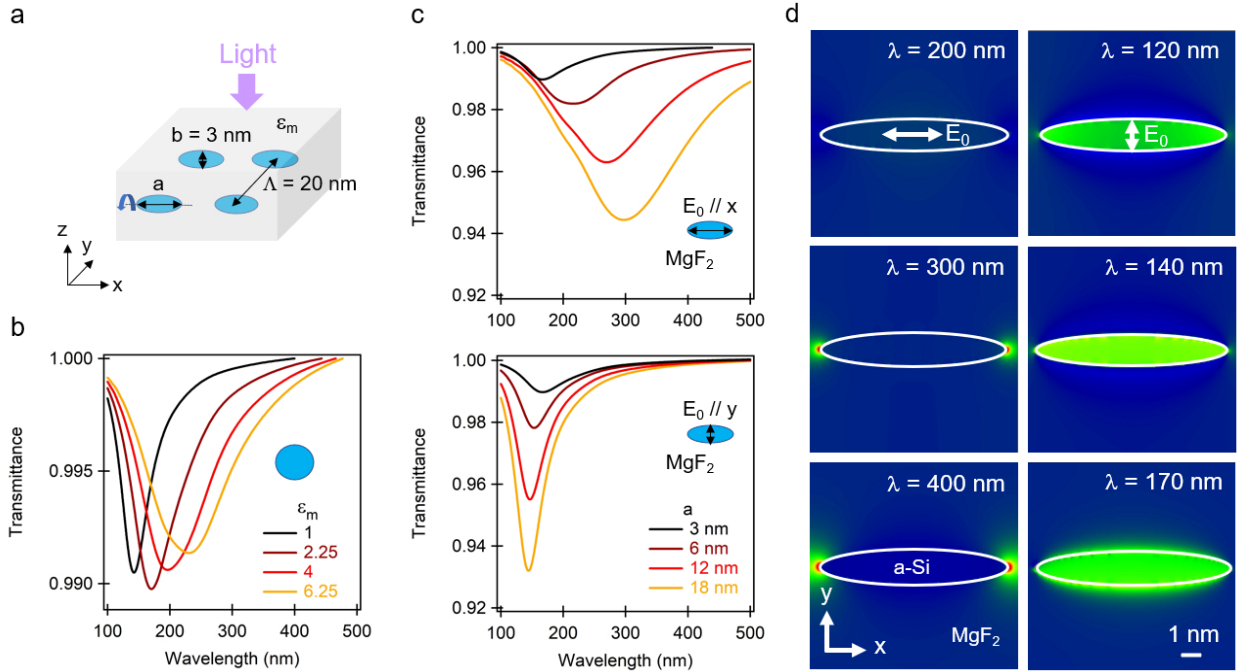

**Figure 4. Ultrathin nanostructures harnessing the full plasmonic spectral region of a-Si.** (a) A planar square array of monodisperse spheroidal a-Si nanostructures embedded in a matrix is considered. Their revolution and long axis a is oriented along the x axis, their short axis is b, and their separation distance is Λ. Light impinges along the z axis, at normal incidence. (b) FDTD-calculated optical transmittance spectra for spherical nanostructures (a = b = 3 nm, Λ = 20 nm) as a function of the dielectric function $\varepsilon_m$ of the fully transparent matrix. (c) FDTD-calculated optical transmittance spectra in the case of prolate spheroidal nanostructures (b = 3 nm, Λ = 20 nm) in a $MgF_2$ matrix, as a function of a, for (top panel) a longitudinal, x-oriented and (bottom panel) a transverse, y-oriented incident electric field $E_0$. (d) FDTD-calculated electric field maps corresponding to the resonances seen in (c) for the nanostructures with a = 18 nm, at the wavelengths of their maximum and half width at half maximum, for a x-oriented $E_0$ (left column) and a y-oriented $E_0$ (right column).

We first considered the case where spherical nanostructures are embedded in UV fully-transparent matrix, for which $\varepsilon_m$ is a real number and was varied between 1 and 6.25. As shown in Figure 4b, the a-Si plasmon resonance is clearly seen without the a-$Al_2O_3$ absorption tail, and, upon decreasing $\varepsilon_m$, this resonance shifts toward shorter wavelengths well into the 100-200 nm wavelength region. Thus, to achieve well-defined plasmonic features in the vacuum UV region, a UV-transparent matrix with a lower $\varepsilon_m$ than a-$Al_2O_3$ is required. Alkaline and alkaline-earth metal fluorides, such as LiF and $MgF_2$, respectively, fulfil these criteria.

We thus performed simulations in which $MgF_2$ was chosen as embedding matrix, and prolate spheroidal nanostructures were specifically oriented with their revolution and long axis aligned with the x direction of the array. Their length a was varied, while their width b was fixed. As shown in Figure 4c, by orienting the polarization of the incident electric field along x or y, the longitudinal or transverse in-plane plasmon mode of the nanostructures is observed, respectively. When the axis ratio a/b turns from 1 to 6, the longitudinal mode shifts from 180 nm to 300 nm, and the transverse mode shifts from 180 to 140 nm. For the longest axis ratio considered (a/b = 6), the two modes are thus well spectrally-separated (300 vs 140 nm), showing the potential of this simple design for polarization and wavelength-selective vacuum UV ultrathin film absorbers.

The broad spectral separation of the two plasmonic modes, one in the mid UV and the other in the vacuum UV, together with the weakly negative $\varepsilon_1$ values in the considered spectral region, imply that different mechanisms are involved in the plasmon excitation. To investigate these mechanisms, we performed FDTD calculations of the electric field maps of nanostructures with a/b = 6, for both orientations of the incident electric field. These calculations were done at the wavelength of the longitudinal (300 nm) and transverse (140 nm) plasmon resonances, and at those corresponding to their half width at half maximum.

From the corresponding maps shown in Figure 4d, it appears that the longitudinal mode at 300 nm leads to an enhanced field at the apex of the nanostructure, at its external surface. This is the fingerprint of a surface plasmon. It results from the negative $\varepsilon_1$ and high $\varepsilon_2$ of a-Si at the wavelength of 300 nm, which both prevent the field to penetrate into the nanostructure. A similar field map is observed at 400 nm, where both the high positive $\varepsilon_1$ and $\varepsilon_2$ of a-Si enable a surface plasmon-like field pattern. At 200 nm, in contrast, a homogeneous weakly enhanced field is seen inside the nanostructure. This is an epsilon-near-zero fingerprint, yet still faint, which appears because a-Si displays a weakly negative $\varepsilon_1$ and a small $\varepsilon_2$ at this wavelength.

For the transverse mode at 140 nm, and at 120 and 170 nm, where a-Si displays a weaklier negative $\varepsilon_1$, a much clearer field enhancement is seen in the nanostructure. This suggests that a so-called epsilon-near-zero plasmon is excited [30-32]. At 170 nm, where $\varepsilon_1$ shows slightly more negative values than at 120 and 140 nm, some field enhancement can in addition be seen at the external surface of the nanostructure, suggesting that a surface plasmon is also excited.

Summarizing, we have both theoretically and experimentally demonstrated that interband plasmonic phenomena take place in ultrathin a-Si nanostructures embedded in a matrix, in the vacuum UV to near UV spectral region. Clear spectral resonances, showing either surface plasmonic or epsilon-near-zero plasmonic features, can be observed down to the vacuum UV provided a suitable matrix, presenting a wide bandgap to ensure transparency in the broadest spectral range and a low dielectric permittivity, is chosen, while the shape and orientation of the nanostructures are adequately tailored. We have proposed an ultrathin film design suitable for achieving polarization and wavelength-selective vacuum UV absorption. It consists of deeply-subwavelength unidirectionally oriented a-Si nanostructures embedded in a UV-transparent $MgF_2$ matrix.

These findings reveal a-Si as a promising material platform for ultracompact plasmonic devices operating down to vacuum UV wavelengths, for applications including anticounterfeiting, data encryption and storage, sensing and detection. The relevance of this material is not only due to its appealing properties. It also builds on the wide earthly abundance of Si, and on the possibility to grow a-Si, using standard deposition methods such as pulsed laser deposition, sputtering, evaporation, or chemical vapor deposition [33-34], in some cases using room temperature processes.

Our study pioneers the demonstration of the suitability of a-Si nanostructures for UV plasmonics. It is noteworthy that this direction has remained unexplored until now, despite the extensive studies on the UV dielectric function and electronic properties of a-Si, which have provided accurate data since the 70s. This is partly due to the fact that efforts to find alternative plasmonic materials beyond Ag and Au, along with their tailored applications, only started around 2010 [1]. In this context, it is particularly illustrative that the UV plasmonic properties of Al nanostructures, a widespread metal, were not studied in depth until 2013 [22-25]. Even more recent is the concept and application of interband plasmonics based on non-metals [14-15, 35-36]. Finally, it is also interesting to note that the Si nanophotonics community, to which our group belongs [28, 37-39], focused its research efforts and resources on investigating the properties and applications of Si nanostructures in the visible and near IR spectral region [28, 37-38, 40-43], where they display tunable light emission and sunlight harvesting properties, while the UV was neglected as no immediate widespread applications in this spectral region were foreseen.

Therefore, our findings open a new unexplored path that may appeal both for the plasmonics and Si-nanophotonic communities. Furthermore, they raise the fundamental question: what is the minimum number of atoms and local order enabling interband plasmonic properties? Previously, such properties were reported in amorphous p-block chalcogenide nanostructures with characteristic sizes in the 100 nm range [44]. Herein, we observed them in 1 order of magnitude smaller, ultrathin a-Si nanostructures. Future works may aim at finding how the local atomic order affects interband plasmonic properties, and whether such properties still exist in few-atom nanostructures.

## Methods

*Fabrication and material characterization.* The films were deposited by pulsed laser deposition following the method described in [28], at room temperature on c-Si and 1 mm-thick $MgF_2$ substrates. Cross-sectional HRTEM characterization was done on a film deposited on c-Si after FIB preparation. XPS characterization was done as described in [28], on a film deposited on c-Si, after steps of ion-beam etching that enabled recording in-depth photoelectron spectra from the film surface to the Si substrate.

*Optical measurements and analysis.* Optical transmittance characterization was done on films grown on $MgF_2$, in the 114-200 nm wavelength range using a home-made vacuum spectrophotometer and in the 200-1500 nm wavelength region using a Varian Cary 5000 dual-beam spectrophotometer. All measurements were done at normal incidence, with unpolarized light.

Spectroscopic ellipsometry characterization was done on films grown on Si, in the 250-1700 nm wavelength range using a Woollam VASE system. Measurements were done at angles of incidence of 65, 70, 75º. Spectra were fitted using the transfer matrix formalism considering each sample as an effective, homogeneous film on a semi-infinite c-Si substrate. By restricting the analysis range to 600-1700 nm, where the films are transparent and their refractive index is wavelength-independent, both their thickness and effective near IR refractive index were obtained: $Si:Al_2O_3$ ($n_{eff}$ = 1.77), t = 18.2 nm; $Si:Al_2O_3$ ($n_{eff}$ = 1.67), t = 16.7 nm; $Al_2O_3$ n = 1.64, t = 14.3 nm).

Transmittance spectra were then fitted in the 114-1500 nm range using the transfer matrix formalism, considering each sample as an effective, homogeneous film on a 1 mm-thick $MgF_2$ substrate, in order to accurately determine the effective dielectric function $\varepsilon_{eff}$ of the films. $\varepsilon_{eff}$ was described by a sum of Kramers Kronig-consistent oscillators with real offset. Film thicknesses were fixed at the values obtained by ellipsometry. The fits were run so that the final $\varepsilon_{eff}$ also enables matching ellipsometry data, in particular that the near-IR $n_{eff}$ value coincides with the ellipsometry-determined one. The dielectric function of $MgF_2$ was determined by combining ellipsometry and transmittance measurements.

*Simulations.* Effective medium simulations were done using a home-made program, based on the realistic and advanced theoretical background that was described and validated in [27] and [29], considering the layered film as a homogeneous effective medium. FDTD simulations were done using the OPTIFDTD software. The nanostructure array was described using periodic boundary conditions in the x-y plane, while perfect matching layers were used in the z direction. An adaptative mesh was used. A pulse with gaussian temporal profile was sent onto the array, and the transmitted wave was recorded in a plane placed under it, and then Fourier processed to obtain the transmittance spectrum.

## Aknowledgements

This work has been partly funded by the national research grants ALPHOMENA (PID2021-123190OB-I00f) and ASSESS (TED2021-129666B-C21 and C22) funded by MCIN/AEI/10.13039/501100011033 and by the European Union NextGenera9onEU/PRTR and the European Regional Development Fund (ERDF), and by the national research grant CoatFuvEspIon (PID2022-142417OB-I00) funded by (MCIN/AEI/10.13039/501100011033, European Union NextGeneration EU/PRTR). Authors acknowledge the use of instrumentation as well as the technical advice provided by the National Facility ELECMI ICTS, node "Laboratorio de Microscopías Avanzadas" at University of Zaragoza.

**Supporting Information**

**S1. Optical transmittance of Si:Al$_2$O$_3$ films including a continuous ultrathin a-Si layer**

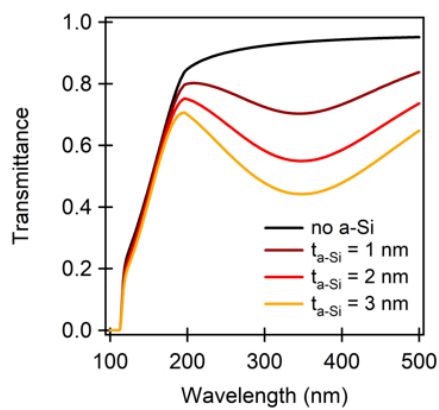

**Figure S1. Simulated optical transmittance spectra of a-Al$_2$O$_3$/a-Si/a-Al$_2$O$_3$ layered films in which the a-Si layer is continuous, on a MgF$_2$ substrate.** The a-Si and a-Al$_2$O$_3$ layer thicknesses are $t_{a-Si}$ and 10 nm, respectively. For each value of $t_{a-Si}$, a band with lowered transmittance centered around 350 nm is seen. Such feature is absent from the measured transmittance spectra of the Si:Al$_2$O$_3$ ($n_{eff}$ = 1.77) and Si:Al$_2$O$_3$ ($n_{eff}$ = 1.67) films (see Figure 3c), suggesting that the a-Si layers in these films are not continuous.

**S2. Evidence of UV plasmon resonances in other Si:Al$_2$O$_3$ films**

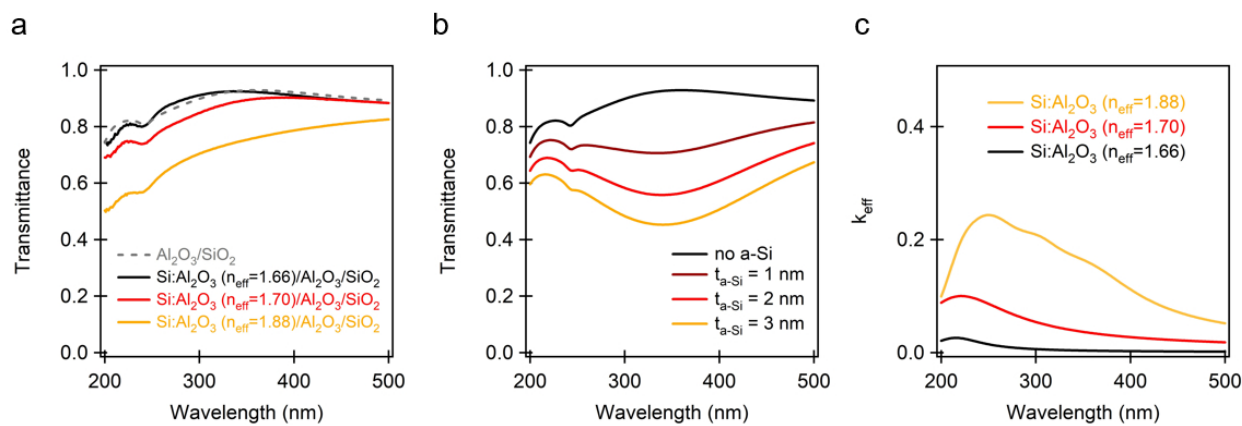

**Figure S2. Optical properties of a-Al$_2$O$_3$/a-Si/a-Al$_2$O$_3$ layered films pulsed-laser deposited on a SiO$_2$ substrate. The films contain different amounts of Si.** The top and bottom Al$_2$O$_3$ layer thicknesses are aproximately 10 nm and in the 70-95 nm range, respectively. They are modeled as a two-layer structure on SiO$_2$, i.e., Si:Al$_2$O$_3$/Al$_2$O$_3$/SiO$_2$, and are labeled with the effective refractive $n_{eff}$ of the Si:Al$_2$O$_3$ layer, which is 21 nm thick. (a) Measured spectra together with that of a reference Al$_2$O$_3$/SiO$_2$ sample. (b) Simulated optical transmittance spectra of a a-Al$_2$O$_3$/a-Si/a-Al$_2$O$_3$/SiO$_2$ layered structure in which the a-Si layer is continuous with a thickness $t_{a-Si}$. The top and bottom Al$_2$O$_3$ layer thicknesses are 10 nm and 95 nm. (c) $k_{eff}$ spectra of the Si:Al$_2$O$_3$ layer for the films shown in (a).